# Tuning Thermal Conductivity of Hybrid Perovskites through Halide Alloying

Guang Wang[1#], Hongzhao Fan[1#], Zhongwei Chen[2#], Yufei Gao[3], Zuankai Wang[4], Zhigang Li[1], Haipeng Lu[2*] and Yanguang Zhou[1*]

[1]*Department of Mechanical and Aerospace Engineering, The Hong Kong University of Science and Technology, Clear Water Bay, Kowloon, Hong Kong SAR*

[2]*Department of Chemistry, The Hong Kong University of Science and Technology, Clear Water Bay, Kowloon, Hong Kong SAR*

[3]*Key Laboratory of Ocean Energy Utilization and Energy Conservation of Ministry of Education, School of Energy and Power Engineering, Dalian University of Technology, Dalian, China*

[4]*Department of Mechanical Engineering, The Hong Kong Polytechnic University, Hong Kong SAR*

**Abstract**

Tuning the thermal transport properties of hybrid halide perovskites is critical for their applications in optoelectronics, thermoelectrics, and photovoltaics. Here, we demonstrate an effective strategy to modulate the thermal transport property of hybrid perovskites by halide alloying. A highly tunable thermal conductivity of mixed-halide hybrid perovskites is achieved due to halide-alloying and structural distortion. Our experimental measurements show that the room temperature thermal conductivity of MAPb(Br$_x$I$_{1-x}$)$_3$ ($x$ = 0–1) can be largely modulated from 0.27±0.07 W/mK ($x$ = 0.5) to 0.47±0.09 W/mK ($x$ = 1). Molecular dynamics simulations further demonstrate that the thermal conductivity reduction of hybrid halide perovskites results from the suppression of the mean free paths of the low-frequency acoustic and optical phonons. It is found that halide alloying and the induced structural distortion can largely increase the scatterings of optical and acoustic phonons, respectively. The confined diffusion of MA$^+$ cations in the octahedra cage is found to act as an additional thermal transport channel in hybrid perovskites and can contribute around 10-20% of the total thermal conductivity. Our findings provide a strategy for tailoring the thermal transport in hybrid halide perovskites which may largely benefit their related applications.

---

[#] These authors contribute equally. [*] Author to whom all correspondence should be addressed. Email: haipenglu@ust.hk and maeygzhou@ust.hk

## INTRODUCTION

Hybrid organic-inorganic metal-halide perovskite has emerged as one of the most intriguing and promising materials due to its excellent photophysical properties, such as superior light adsorption (*1*, *2*), tunable optical bandgaps (*3*), long carrier diffusion length (*4*), high charge carrier mobility and enhanced carrier lifetimes (*5–7*). These advantages benefit their applications in photovoltaics (*3*, *8–10*), optoelectronics (*11–14*), lasers (*15*, *16*), and thermoelectrics (*17*). Some halide perovskite-based devices such as photovoltaics and LEDs have thermal stability issues when the working temperature is too high, which is attributed to the ultralow thermal conductivity of the halide perovskite (i.e., < 1 W/mK) (*18*). Meanwhile, halide perovskite is also known as a "phonon glass, electron crystal", making it a promising candidate for thermoelectrics (*19*) that require low thermal conductivity but high electrical properties. Understanding the dynamics of thermal transport and heat carriers in halide perovskites is therefore crucial for guiding the design of these devices. For example, the lifetime of photoexcited carriers is strongly related to the electron-phonon scattering (*20*), and shorter phonon mean free path can have significant effects on the hot carriers cooling and recombination in solar cells (*21*).

Mixed-halide hybrid perovskite (MHHP) such as $MAPb(X_xY_{1-x})_3$ (MA = $CH_3NH_3^+$; X and Y = $Cl^-$, $Br^-$, $I^-$; $x$ = 0-1) has been widely used for optoelectronics owing to its tunable bandgap (i.e., 1.53 to 2.9 eV), with corresponding adsorption spectrums ranging from 420 to 836 nm (*19*). However, the thermal transport properties of MHHP, which strongly influence the performance of MHHP-based optoelectronics, become nontrivial. The crystal structure of hybrid perovskite ($MAPbX_3$) consists of 3D corner-sharing $[PbX_6]^{4-}$ octahedra, with the $MA^+$ cations occupying voids within the cage formed by the $PbX_6$ octahedra framework. The highly anharmonic motion of $MA^+$ cations in $MAPbX_3$ induced by the soft and flexible inorganic framework (*22*, *23*) will result in a low thermal conductivity of ~0.4 W/mK (*24*). For a mixed-halide perovskite structure, a different halide ion is introduced into the inorganic framework, forming the alloyed structure, namely, MAPb($X_xY_{1-x}$)$_3$. It is known that the thermal conductivity of a semiconductor alloy is generally much lower than that of its crystalline counterpart due to the mass disorder (25) and structural discontinuity (26). Therefore, we hypothesize that a reduced thermal conductivity would be obtained in the mixed-halide perovskite. In alloyed all-inorganic halide perovskite, the dynamic cation off-centering can also induce ultralow thermal conductivity (27). However, in the hybrid halide perovskite, the diffusion of $MA^+$ cations in the octahedra cage of perovskites may introduce an extra thermal transfer channel, which benefits the thermal energy exchange(23). The alloyed structure of MHHPs introduces asymmetric interactions between cations and anions and promotes the diffusion of $MA^+$ cations, which may increase the thermal conductivity contributed by the diffusion of $MA^+$ cations. An intrinsic question is then raised: what is the lower limit and tunable range of the thermal conductivity of MHHPs?

In this paper, we systematically investigate the thermal transport properties and dynamics of heat carriers of MAPb(Br$_x$I$_{1-x}$)$_3$ ($x$ = 0-1) at room temperature using both the frequency-domain thermoreflectance (FDTR) measurements and molecular dynamics (MD) simulations. Our FDTR results show that the minimal thermal conductivity of MAPb(Br$_x$I$_{1-x}$)$_3$ can be achieved is 0.27±0.07 W/mK, where $x$ is equal to 0.5. The thermal conductivity of MAPb(Br$_x$I$_{1-x}$)$_3$ ($x$ = 0-1) can be largely modulated from 0.27±0.07 W/mK ($x$=0.5) to 0.47±0.09 W/mK ($x$=1). The minimal thermal conductivity is reduced by 42.5% and 22.8% compared to that of MAPbBr$_3$ and MAPbI$_3$, respectively. Our MD simulations demonstrate that the reduction of thermal conductivity of MMHP is mainly attributed to the suppression of the mean free paths (MFPs) of the low-frequency acoustic and optical phonons, which results from the strong phonon scatterings caused by the alteration of local potential landscape and alloying. Furthermore, the thermal conductivity contribution resulting from the diffusion of $MA^+$ cations is found to be non-negligible, which increases from ~10% (MAPbI$_3$ and MAPbBr$_3$) to ~20% (MAPb(Br$_{0.5}$I$_{0.5}$)$_3$). Our findings provide new insights into thermal transport in MHHP, which will facilitate its applications in optoelectronics, thermoelectrics, and photovoltaics.

**RESULTS**

**Materials synthesis and characterization**. The structure of MAPb(Br$_x$I$_{1-x}$)$_3$ is similar to that of MAPbBr$_3$ and MAPbI$_3$, of which the Br and I ions are randomly distributed on the halide sites. It is noted that the MAPbI$_3$ and MAPbBr$_3$ are tetragonal and cubic structures at room temperature (**Figure 1a**), respectively. Therefore, the superstructure of MAPb(Br$_x$I$_{1-x}$)$_3$ depends on the composition ratio between Br and I ions. Here, high-quality MAPb(Br$_x$I$_{1-x}$)$_3$ crystals with grain sizes of ~ several millimeters are synthesized following the method reported in Ref. (*28*) (see **Methods** for details). **Figure 1b** is the digital photos of the as-synthesized MAPb(Br$_x$I$_{1-x}$)$_3$ crystals in which $x$ is 0, 0.17, 0.5, 0.75, and 1. The composition of the MHHPs can be distinguished by their color, which changes from dark black to orange with the higher ratio of the Br component. The ratio of the elements in MHHPs is measured by inductively coupled plasma mass spectrometry (ICP-MS), and the results slightly deviate from the stoichiometric ratios (see **Supplementary Note 1** for details). All the ratios of halide ions in MHHPs hereafter refer to the stoichiometric ratios. We also grind the MHHPs into small pieces for the high-resolution transmission electron microscopy (HRTEM) measurement (**Figure 1c**). The clear lattice patterns and corresponding fast Fourier transform pattern images (the inset of **Figure 1c**) demonstrate the single-crystalline nature of the MHHPs. Powder x-ray diffraction (XRD) data of MAPb(Br$_x$I$_{1-x}$)$_3$ ($x$ = 0, 0.17, 0.5, 0.75, 1) indicate a high phase purity as all the peaks can be easily indexed based on the parent structure of halide perovskites (**Figure 1d**). The diffraction patterns show that MAPbI$_3$ crystallizes a tetragonal phase, and MAPbBr$_3$ is in a cubic phase at room temperature (**Figure S2**). A phase transition is observed when the Br ratio changes from 0 to 0.5, as indicated by the magnified (1 0 0)$_c$ and (2 0 0)$_c$ peaks. Meanwhile, the MHHP may be treated as a pseudo cubic phase when $x > 0.21$, where cubic and tetragonal phases co-existed and the cubic phase dominated (*19*).

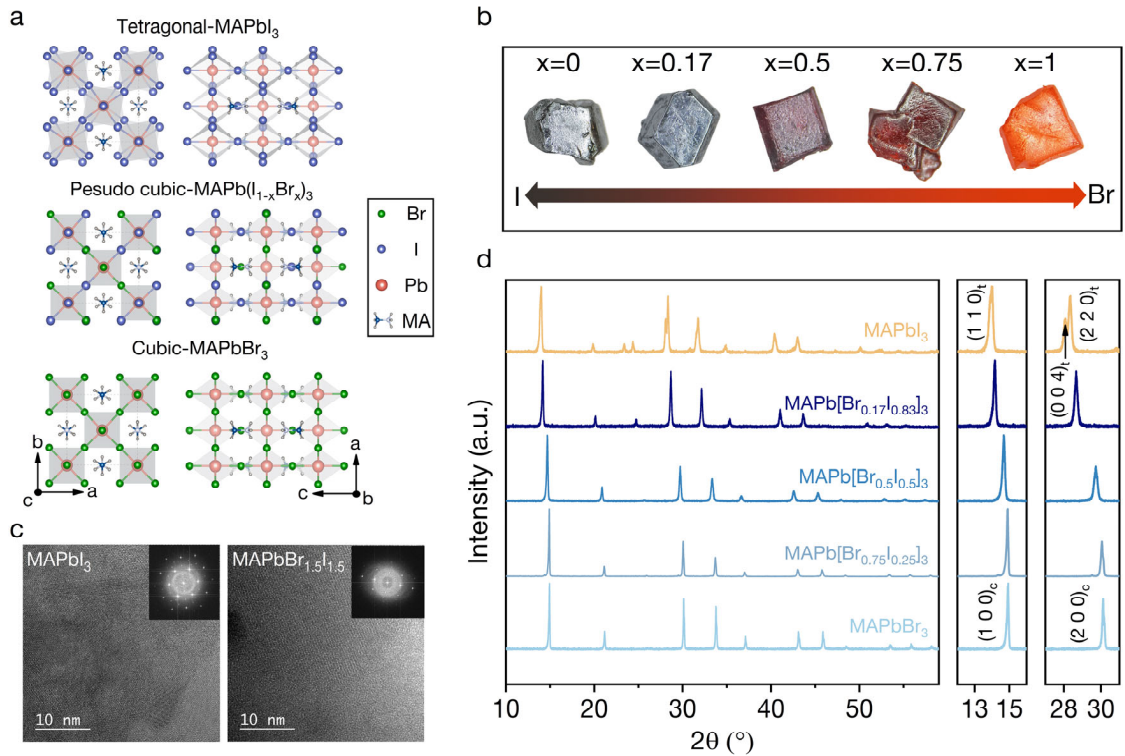

**Figure 1 Characterization of MHHPs.** a. The crystal structure of MHPs. b. The digital photos of synthesized MHPs series MAPb(Br$_x$I$_{1-x}$)$_3$ ($x$ = 0, 0.17, 0.5, 0.75, 1). c. The HRTEM images of MHHPs pieces and the corresponding fast Fourier transform spots of MAPbI$_3$ and MAPb(Br$_{0.5}$I$_{0.5}$)$_3$ show the single-crystalline nature. d. The powder *x*-ray diffraction (PXRD) patterns of the MAPb(Br$_x$I$_{1-x}$)$_3$ ($x$ = 0, 0.17, 0.5, 0.75, 1) indicate the phase evolution from cubic to tetragonal phase.

**FDTR measurements**. The thermal transport properties of crystalline MHHPs were then characterized using optical pump-probe spectroscopy based on the FDTR (*29, 30*). We first finely polished these crystals with irregular shapes and non-flat surfaces to ensure a good thermoreflectance signal that can be detected in our FDTR experiments (see **Methods** for details). A ~100 nm Au film was then sputtered on the surfaces of the crystals as a transducer layer (see **Supplementary Note 2**), which could generate a rapid temperature rise once the pump laser irradiated on the surface. The phase lag between the pump laser and the probe laser was determined by a lock-in amplifier and fitted using a heat diffusion model (*31*, *32*). The intensity radii of the pump and probe laser were acquired by a beam offset method and fitted to a Gaussian profile each time before the phase lag measurement, which was ~3.6 μm and ~5 μm, respectively. A 10X optical microscope was used to find the regions with smooth surfaces and high signal-to-noise ratios, which were critical for the FDTR measurements.

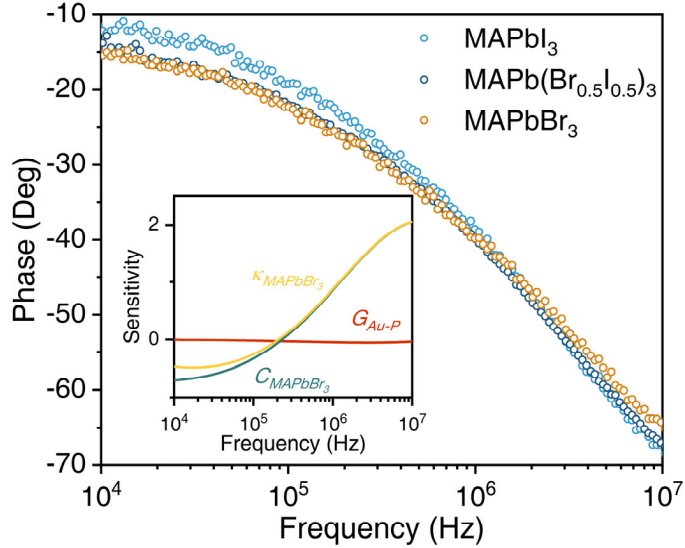

**Figure 2 Thermal conductivity measurement of MHHPs by FDTR**. The representative FDTR signal phase as a function of pump beam modulation frequency of three typical samples: $MAPbI_3$, $MAPb(Br_{0.5}I_{0.5})_3$, and $MAPbBr_3$ at room temperature. The inset picture is the sensitivity analysis of $MAPbBr_3$.

**Figure 2** shows the representative phases of $MAPbI_3$, $MAPb(Br_{0.5}I_{0.5})_3$, and $MAPbBr_3$ at room temperature measured by FDTR. Our measured phase lags indicate that the thermal conductivity of $MAPb(Br_{0.5}I_{0.5})_3$ is lower than that of $MAPbI_3$ and $MAPbBr_3$, of which the phase lag is larger than the other two across the modulation frequency range. We analyzed the sensitivity of the related parameters included in our measurement, as shown in the inset of **Figure 2**. We found that the sensitivities of the thermal conductivity and heat capacity of $MAPbBr_3$ have a large overlap in the frequency range we focus on. Therefore, for $MAPbBr_3$, the thermal conductivity and heat capacity are dependent parameters in the FDTR fitting model.

In our experiments, we only measure the thermal conductivity. The heat capacity of all systems with various composition ratios is calculated based on the values from reference(*18*) (see **Supplementary Note 3**). The same spot was swept 5 times for each sample to get the average thermal conductivity. For each case, the measurements were conducted in more than 5 different samples, and several spots in one sample were measured. The statistical mean value is calculated by Gaussian fitting of all measured results (**Figure S5**).

The thermal conductivities of crystalline MAPbBr$_3$ and MAPbI$_3$ at room temperature are determined as 0.47±0.09 W/mK and 0.35±0.11 W/mK, respectively, which agree well with the results from the reference (*18*). A minimal thermal conductivity of 0.27±0.07 W/mK is found for MAPb(Br$_{0.5}$I$_{0.5}$)$_3$ at room temperature. The room temperature thermal conductivities for the other two alloyed halide perovskites, i.e., MAPb(Br$_{0.17}$I$_{0.83}$)$_3$ and MAPb(Br$_{0.75}$I$_{0.25}$)$_3$, are determined as 0.33±0.06 W/mK and 0.44±0.17 W/mK, respectively (**Figure 3a**). Our FDTR measurements here show that the thermal conductivity of MAPb(Br$_x$I$_{1-x}$)$_3$ firstly decreases and then increases with the ratio of Br ($x$). The minimal thermal conductivity of MHHPs MAPb(Br$_x$I$_{1-x}$)$_3$ is reduced by 42.6% (22.9%) compared to that of MAPbBr$_3$ (MAPbI$_3$). It indicates that the alloying of halide atoms in MHHPs is an effective approach to modulating the corresponding thermal transport properties. Furthermore, we would like to emphasize that the size of our samples in our FDTR measurements is several millimeters, and therefore the size effect on the measured thermal conductivity can be ignored as the corresponding vibrational mean free paths (MFPs) are smaller than 10 nm (see analysis below).

**Two-channel thermal transport in MHHPs.** To uncover the underlying mechanisms behind the thermal transport in MHHPs, we performed MD simulations to calculate the thermal conductivity and the phonon information such as the mean free path of MHHPs. While the thermal conductivity of MHHPs calculated using MD simulations slightly differs from the experimental measurements, our simulation results show the same trend as the FDTR measurements (**Figure 3a**). Both MD simulations and experimental measurements find that the thermal conductivity of MHHPs firstly decreases and then increases with the ratio of Br and reaches a minimal value when the ratio of Br is 0.5 (**Figure 3a**).

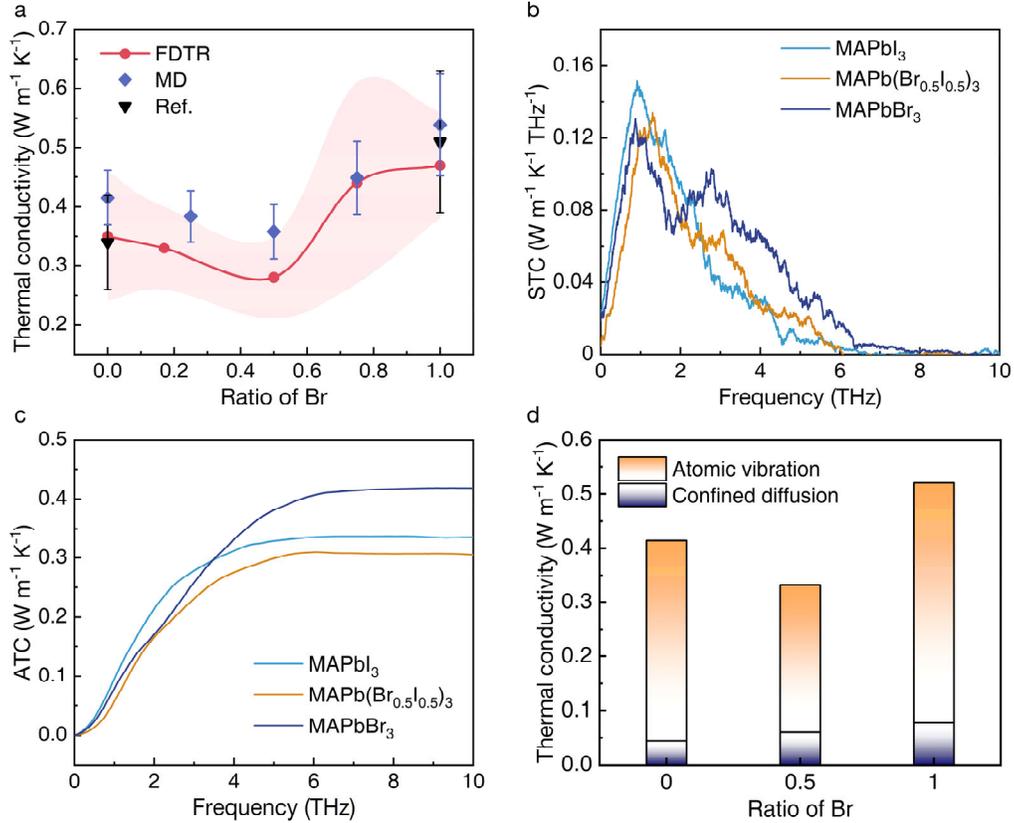

**Figure 3 Characterization of thermal transport in MHHPs.** a. The thermal conductivity of MHPs from experimental measurements (red line, in which the pink shadow area is the stand errors of FDTR measured results), molecular dynamics simulation (blue dot), and references (black dot) (*18*). b. The spectral thermal conductivity contributed by atomic vibrations in MAPbI$_3$, MAPb(Br$_{0.5}$I$_{0.5}$)$_3$, and MAPbBr$_3$. c. The accumulative thermal conductivity of MHPs. d. The decomposed contribution from atomic vibrations and confined diffusion to total thermal conductivity.

We further calculate the spectral thermal conductivity (STC) of three typical MHHPs, i.e., MAPbI$_3$, MAPb(Br$_{0.5}$I$_{0.5}$)$_3$ and MAPbBr$_3$, using our frequency direct domain decomposed method (FDDDM) (*33–35*), which can quantitatively characterize the thermal transport in MHHPs. Our calculated results show that the thermal conductivity of all three MHHPs is mainly contributed by vibrations with frequencies smaller than 6 THz (**Figure 3b**). For MAPbI$_3$, the acoustic phonons (i.e., their vibrational frequencies are smaller than 2 THz) contribute ~60% to the total thermal conductivity (**Figures 3c** and **Figure 4**). The rest ~40% is contributed by the optical phonons with frequencies ranging from 2 to 6 THz. For MAPbBr$_3$, it is interesting to find that the optical phonons with frequencies of 2~6 THz contribute ~62% to the total thermal conductivity (**Figures 3c** and **Figure 4**). These acoustic phonons with frequencies smaller than 2 THz only contribute around 38% to the total thermal conductivity. This is because the intrinsic scattering among phonons in MAPbBr$_3$ is much stronger than that in MAPbI$_3$, as discussed below (**Figure 4**). Meanwhile, the group velocities of these optical phonons in MAPbBr$_3$ are generally larger than those in MAPbI$_3$. Consequently, while the contribution to the total thermal conductivity from acoustic phonons in MAPbBr$_3$ is smaller than that in MAPbI$_3$, the total thermal conductivity of MAPbBr$_3$ is much larger than that of MAPbI$_3$ (**Figure 3c**). For MAPb(Br$_{0.5}$I$_{0.5}$)$_3$, we find that the modal thermal conductivity of acoustic phonons (i.e., frequencies smaller than 2 THz) is generally smaller than that in MAPbI$_3$ and the spectral thermal conductivity of optical phonons are smaller than that in MAPbBr$_3$ (**Figure 3b**). As a result, the thermal conductivity of MAPb(Br$_{0.5}$I$_{0.5}$)$_3$ decreases compared to that of both MAPbI$_3$ and MAPbBr$_3$.

It is also interesting to note that the total thermal conductivity accumulated based on calculated STC is smaller than that computed using direct non-equilibrium MD simulations (NEMD, **Figures 3a** and **3c**). In our FDDDM calculations, the heat current contributed by the diffusion of ions is ignored and therefore our calculated STC only considers the virial heat current or equivalently the heat current mainly contributed by the lattice vibrations (see **Methods for details**). It is known that the large movement such as the rotation of organic cations in the cage will largely affect the corresponding thermal transport properties of inorganic-organic hybrid perovskites (*36–38*). Some previous simulations (*24, 37, 39*) argued that the low thermal conductivity of MAPbI$_3$ stemmed from the strong scatterings between phonons and rotors of organic cations. Some other studies (*23*, *40*, *41*), on the contrary, suggested that the large movements of organic cations in the cage of MAPbI$_3$ can benefit thermal transport. We then calculate the thermal conductivity contributed by the diffusion of ions (i.e., mainly MA$^+$). The calculated thermal conductivity considering both the lattice vibrations and the confined diffusion of MA$^+$ is equal to the value calculated using direct NEMD simulations. Our results show that the thermal conductivity caused by the confined diffusion of MA$^+$ cannot be ignored and can contribute around 12% to the total value for MAPbI$_3$ and MAPbBr$_3$ (**Figure 3d**). For MAPb(Br$_{0.5}$I$_{0.5}$)$_3$, thermal conductivity contributed from the cation diffusion is found to slightly increase compared to MAPbI$_3$. The contribution to the total thermal conductivity resulting from the diffusion of MA$^+$ cations in MAPb(Br$_{0.5}$I$_{0.5}$)$_3$ is increased to ~20% as the atomic vibrations in MAPb(Br$_{0.5}$I$_{0.5}$)$_3$ are strongly scattered. The reduction of total thermal conductivity for the MHHPs compared to that of MAPbI$_3$ and MAPbBr$_3$ is therefore resulting from the competition between the heat transfer channel of lattice vibrations and the thermal pathway contributed by the diffusion of MA$^+$ cations.

**Underlying mechanisms behind the phonon transport in MHHPs.** To characterize the phonon thermal transport in MHHPs, we calculate the phenomenological mean free paths (MFPs) of phonons. For MAPbI$_3$ and MAPbBr$_3$, the vibrational MFPs are generally smaller than 10 nm, which leads to their low thermal conductivities as discussed above. It is noted that the vibrational MFPs in MAPbI$_3$ are in principle larger than those in MAPbBr$_3$ (**Figure 4a**). As shown in **Figure 3a**, the thermal conductivity of MAPbBr$_3$ is larger than that of MAPbI$_3$ which stems from the higher heat capacity (1.45 MJ/(m$^3$K) for MAPbBr$_3$ and 1.28 MJ/(m$^3$K) for MAPbI$_3$) and larger vibrational group velocity of MAPbBr$_3$ (i.e., the mean group velocity near Γ point is 16.64 Å/ps for MAPbI$_3$ and 20.45 Å/ps for MAPbBr$_3$). For MAPb(Br$_{0.5}$I$_{0.5}$)$_3$, these acoustic vibrations with frequencies smaller than 1 THz are scattered strongly, and thus possess much shorter MFPs compared to the corresponding vibrations in MAPbI$_3$ and MAPbBr$_3$ (**Figure 4a**). The maximum phonon MFP in MAPb(Br$_{0.5}$I$_{0.5}$)$_3$ is ~4 nm which is much shorter than ~10 nm for MAPbI$_3$ and ~7 nm for MAPbBr$_3$ (**Figures 4a** and **4b**). It is known that the introduced alloyed atoms or ions can strongly scatter the high-frequency phonons (*42*). This should be the reason that the MFPs of the optical phonons with frequencies of 2~5 THz in MAPb(Br$_{0.5}$I$_{0.5}$)$_3$ are generally shorter than that of the phonons in MAPbI$_3$ (**Figure 4a**). However, we also note that the low frequency (i.e., < 2 THz) acoustic phonons in MAPb(Br$_{0.5}$I$_{0.5}$)$_3$ are significantly scattered by alloying (**Figures 4a** and **4b**).

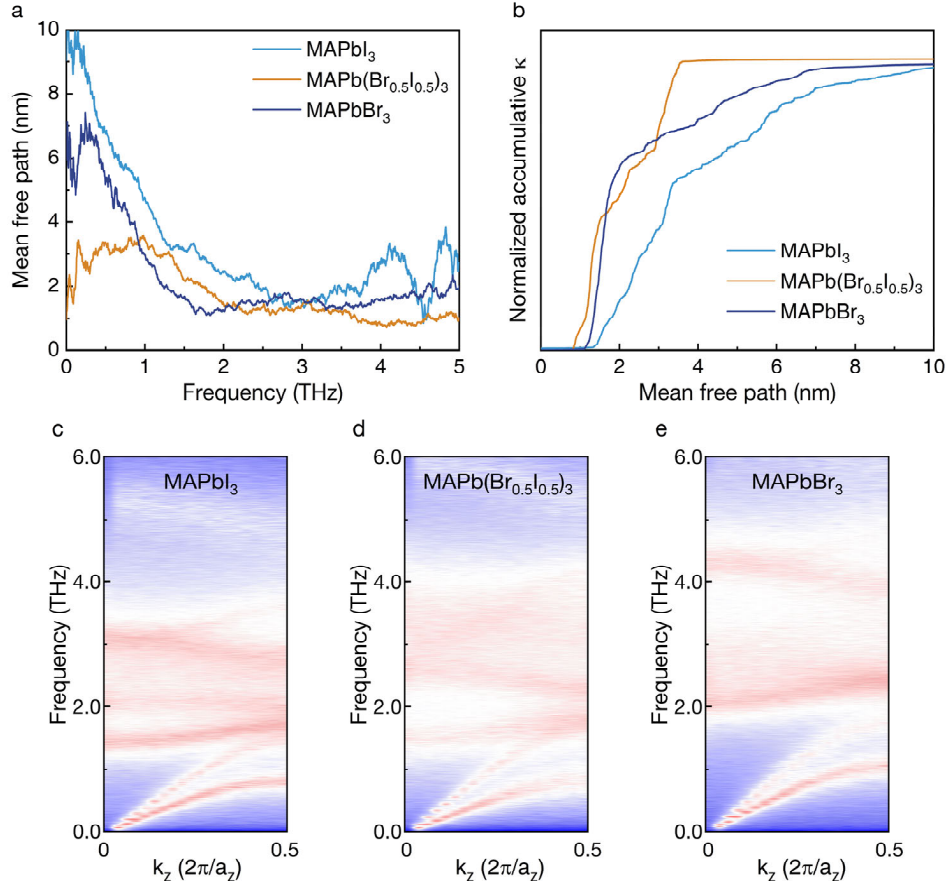

**Figure 4 The phonon transport properties of MHHPs.** a. The frequency-dependent mean free path of MHHPs calculated by MD simulations. b. The normalized accumulative thermal

conductivity of MHPs. c-e. The SED image of MAPbI$_3$, MAPb(I$_{0.5}$Br$_{0.5}$)$_3$, and MAPbBr$_3$, respectively.

We next calculate the spectral energy densities (SEDs) of MAPbI$_3$, MAPb(Br$_{0.5}$I$_{0.5}$)$_3$ and MAPbBr$_3$ (see **Method** for details). Our results show that the scatterings among both acoustic and optical phonons become stronger when alloy is introduced (**Figures 4c-e**). Consequently, the phonon MFPs in MAPb(Br$_{0.5}$I$_{0.5}$)$_3$ are generally shorter than that in MAPbI$_3$ and MAPbBr$_3$ as shown in **Figures 4a** and **4b**, which results in a lower conductive thermal conductivity of MAPb(Br$_{0.5}$I$_{0.5}$)$_3$ (**Figure 3d**). We also find that the scattering of both acoustic and optical phonons in MAPbBr$_3$ is stronger than that in MAPbI$_3$ (**Figures 4c** and **4e**). Therefore, the phonon MFPs in MAPbBr$_3$ are substantially shorter than those in MAPbI$_3$ (**Figure 4a**). However, the heat capacity and group velocity for these phonons in MAPbBr$_3$ become larger compared to that in MAPbI$_3$. This explains that the conductive thermal conductivity of MAPbBr$_3$ is higher than that of MAPbI$_3$ (**Figure 3d**). The alloying of halide atoms in MAPb(Br$_x$I$_{1-x}$)$_3$ will bring two effects, i.e., the mass difference among halide atoms and the alteration of the local potential landscape. It is known that the mass difference in alloys will scatter the high-frequency phonons and then reduce their contribution to the total thermal conductivity (*42*). However, the reduction of thermal conductivity of MAPb(Br$_x$I$_{1-x}$)$_3$ is much smaller than the thermal conductivity decrease observed in experiments when only the mass difference among halide atoms is considered (**see Supplementary Note 4 for details**). It is noted that the structures of MAPbBr$_3$ and MAPbI$_3$ at room temperature are different. This structural difference can cause the alteration of the local potential landscape, which leads to strong anharmonic scatterings among acoustic phonons (**Figure 4d**) and further decreases the thermal conductivity of MAPb(Br$_x$I$_{1-x}$)$_3$. The thermal conductivities of MAPb(Br$_x$I$_{1-x}$)$_3$ are then close to experimental results (*25*, *43*) when both effects are considered.

**CONCLUSIONS**

In summary, the dynamics of thermal transport and heat carriers in MAPb(Br$_x$I$_{1-x}$)$_3$ ($x$ = 0-1) at room temperature are systematically investigated using both the frequency-domain thermoreflectance measurements and molecular dynamics simulations. Our experiments show that the minimal thermal conductivity of MAPb(Br$_x$I$_{1-x}$)$_3$ at room temperature is 0.27±0.07 W/mK when $x$ is equal to 0.5, which is 42.5% and 22.8% lower compared to that of MAPbBr$_3$ and MAPbI$_3$, respectively. By controlling the ratio of halide ions in MAPb(Br$_x$I$_{1-x}$)$_3$ ($x$ = 0-1), the corresponding thermal conductivity can be broadly tuned by 2 times, i.e., from 0.27±0.07 W/mK to 0.47±0.09 W/mK. Our molecular dynamics simulations further find that the reduction of thermal conductivity of MAPb(Br$_x$I$_{1-x}$)$_3$ (0 < $x$ < 1) results from the suppression of the mean free paths of the low-frequency acoustic and optical phonons. This is because of the strong phonon scatterings caused by the alteration of the local potential landscape and alloying when the halide ions are alloyed. Furthermore, the confined diffusion of MA$^+$ cations in the octahedra cage is found to contribute ~10% of the total thermal conductivity for MAPbI$_3$ and MAPbBr$_3$. This contribution increases to ~20% for MAPb(Br$_{0.5}$I$_{0.5}$)$_3$, which stems from the more effective thermal transport of MA$^+$ cation diffusions. Our work here provides new insights into the dynamics of thermal transport in mixed hybrid halide perovskites, which may promote their optoelectronic, thermoelectric, and photovoltaic applications.

**Data and code availability**

The data and code that support the findings of this study are available from the corresponding author upon request.

## METHODS

**Synthesis of MHHPs**

All the chemicals including methylammonium hydrobromide (MABr, 99%, Lumtec), PbBr$_2$ (99.9%, Macklin), PbI$_2$ (99.9%, Aladdin), methylammonium iodide (MAI, 99.5%), Dimethylformamide (DMF, 99.8%, Macklin), γ-butyrolactone (GBL, 99%, Macklin), Poly (propylene glycol) (PPG-3000, molecular weight 3000 Da, Aladdin), were used to synthesize MHHPs as received without further purification. For the synthesis of MAPbBr$_3$ single crystals, 2 mmol PbBr$_2$ and MABr were dissolved in 2 mL DMF and stirred at room temperature for 4h, the solution was filtered using a PTFE filter with 0.2 μm pore size, the filtered solution was then sealed in a vial and kept in an oil bath at 90 °C for crystallization. The MAPbI$_3$ and mixed halide perovskites MAPb(Br$_x$I$_{1-x}$)$_3$ were prepared according to Ref. (*28*). The perovskite precursors MABr, PbBr$_2$, MAI, and PbI$_2$ with stoichiometric ratios were dissolved in 2 mL GBL/DMF. The solution was then stirred at room temperature for 4h. A certain amount of PPG-3000 polymers was also added to the solution to control the nucleation process of MHHPs. The precursor solution was finally filtered using a PTFE filter with a 0.2 μm pore size. The filtered solution was sealed in a vial and kept in an oil bath at 90-95°C until the desired MHHP crystals were formed. The details for synthesizing the MHHPs are given in **Table 1**.

Table 1. Detailed conditions for the synthesis of MHHP crystals.

| | The precursors' concentration | Solvent | The concentration of PPG-3000 (g/mL) | Temperature (°C) |
|---|---|---|---|---|
| MAPbI$_3$ | 1.3 M | GBL | 0.03 | 95 |
| MAPb(Br$_{0.17}$I$_{0.83}$)$_3$ | 1.3 M | GBL/DMF ($v$:$v$ = 1:1) | 0.02 | 90 |
| MAPb(Br$_{0.5}$I$_{0.5}$)$_3$ | 1.3 M | GBL/DMF ($v$:$v$ = 1:1) | 0.1 | 90 |
| MAPb(Br$_{0.75}$I$_{0.25}$)$_3$ | 0.7 M | GBL/DMF ($v$:$v$ = 1:1) | 0.1 | 90 |
| MAPbBr$_3$ | The synthesis method for MAPbBr$_3$ single crystal is different from other perovskites, which has been described above in detail. | | | |

**Characterization of MHHPs**

The powder XRD data of MHHPs was collected on the PANalytical Aeris powder-X-ray diffractometer with a Cu $K_{\alpha 1}/K_{\alpha 2}$ source ($\lambda$ = 1.54051/1.54433 Å). To obtain the high-resolution lattice images, the MHHP crystals with large sizes were first grounded in an agate mortar and then dispersed in Ether. The suspension was then dropped on the TEM grid. The lattice pattern of the thin pieces obtained can be observed using scanning transmission electron microscopy (STEM, JEM-ARM200F JEOL). The thermogravimetric analysis (TGA) was measured from room temperature to 800 °C with a ramp rate of 10 °C/min in a Nitrogen environment (Discovery TGA5500, TA). The electrical properties of the MHHP crystals were measured using a two-probe method (Keithley 2450 resistance measurement system), in which the silver paste was coated on two ends of the MHHP crystals as electrodes. Meanwhile, to validate the composition ratio of as-synthesized MHHPs, we also did the element analysis using ICP-MS (Agilent 7800). The MHHP crystals were dissolved in a mixture of dilute hydrochloric acid and dilute nitric acid. The element ratios of Pb, Br, and I in $MAPb(Br_{0.75}I_{0.25})_3$, $MAPb(Br_{0.5}I_{0.5})_3$ and $MAPb(Br_{0.17}I_{0.83})_3$ were then determined (see **Supplementary Note 1** for details).

**FDTR measurements**

The FDTR, which is a well-established pump-probe thermal properties measurement apparatus, measures the thermal conductivity of MHHPs. As the as-synthesized MHHP crystals may not have smooth surfaces for a good thermoreflectance signal, we polished the MHHP crystal surfaces using dry mechanical polishing before FDTR measurements. The MHHP crystals were first put on the plastic molds. The molds were then filled with epoxy resin. After solidification, these MHHP crystals were embedded in the epoxy resin. Sandpapers with various meshes, #600, #1200, #2000, and #3000, were applied for the rough polishing, followed by the fine polishing using damping polishing cloths with $Al_2O_3$ powder (0.3 $\mu$m, 0.05 $\mu$m, and without $Al_2O_3$ powder). The compressed air is eventually used to clean the surfaces, as shown in **Figure S4**. The samples are then coated with Au film with a thickness of ~100 nm through magnetron sputtering. The coated Au film serves as an optical transducer to absorb the pump laser, and thus the probe laser can detect the induced temperature rise. For each FDTR test, the radius of the pump laser was measured using a beam offset method. The radii for the pump laser and probe laser in our FDTR apparatus were 3.6 $\mu$m and 5 $\mu$m under a 10× objective, respectively. A lock-in amplifier (HF2LI, Zurich) was used to obtain the phase lag between the pump and probe lasers. The phase lag was then fitted using a heat diffusion model to obtain the thermal transport properties. In our FDTR measurements, we can only fit the thermal conductivity of MHHPs based on the sensitivity analysis (**Supplementary Note 3**). Each spot with a smooth and flat surface under the microscope was swept 5 times to reduce the noise.

**Molecular dynamics simulation**

The thermal conductivity of MHHPs was calculated using the equilibrium molecular dynamics (EMD) simulations method using the LAMMPS package (*44*). The size of the simulation model of EMD simulation is the 16×16×16 supercells. The force fields developed for lead halide hybrid perovskites were used to describe interatomic interactions in MHHPs (*45*, *46*). These force fields have been demonstrated to produce experimentally consistent structure parameters for lead halide hybrid perovskites (*45*, *46*). In all MD simulations, the timestep was 0.5 fs. The particle-particle particle-mesh solver with a relative error of $10^{-5}$ was used to consider the long-range Coulombic interactions. The cutoff of pair interactions is 12 Å. The system was first relaxed in the isothermal-isobaric ensemble (*NPT*) at 0 bar and 300 K for 500 ps. After the systems reached the equilibrium state, the ensemble was switched to the canonical ensemble (*NVT*) for another 500 ps. Then, the simulations run in the the microcanonical ensemble (*NVE*) for aonther 1 ns to perform EMD simulations. The thermal conductivity is calculated by

$$\kappa = \frac{V}{3k_B T^2} \int_0^\infty \langle \mathbf{J}(0) \cdot \mathbf{J}(t) \rangle \mathrm{d}t \tag{1}$$

where $V$ is the system volume, $k_B$ is the Boltzmann constant, $T$ is the simulation temperature, $\mathbf{J}(t)$ is the instanous heat flux. The correlation time in our thermal conductivity calculation is 30 ps. For each case, twelve independent calculations are performed to obtain the stable thermal conductivity.

**Heat carriers' quantifications**

To quantitatively characterize the thermal transport in MHHPs, the spectral thermal conductivity and transmission coefficient were calculated by the FDDDM (*33*, *34*, *47*, *48*). The FDDDM decomposition is implemented in the framework of NEMD simulations. In the NEMD simulations, the heat current transferred across an imaginary interface can be calculated by

$$Q_{left \to right} = \sum_{i \in left} \sum_{j \in right} \left\langle \frac{\partial U_j}{\partial \vec{r}_i} \cdot \vec{v}_i - \frac{\partial U_i}{\partial \vec{r}_j} \cdot \vec{v}_j \right\rangle \tag{2}$$

where $U$ represents the potential energy, $\vec{v}_i$ is atomic velocity and $\vec{r}_i$ is atomic position. The atomic velocity and position were recorded during the NEMD simulation for 1 million steps. Then, the spectral heat current across the imaginary interface can be obtained via

$$Q(\omega) = \mathrm{Re} \sum_{i \in left} \sum_{j \in right} \int_{-\infty}^{+\infty} \left\langle \left. \frac{\partial U_j}{\partial \vec{r}_i} \right|_\tau \cdot \vec{v}_i(0) - \left. \frac{\partial U_i}{\partial \vec{r}_j} \right|_\tau \cdot \vec{v}_j(0) \right\rangle e^{i\omega\tau} d\tau \tag{3}$$

The frequency-dependent thermal conductivity is further calculated by

$$\kappa(\omega) = \frac{Q(\omega)}{A \cdot \nabla T} \tag{4}$$

Meanwhile, the generalized vibrational transmission function in NEMD simulations can be calculated as

$$\mathrm{T}(\omega) = \frac{Q(\omega)}{k_B \Delta T} \tag{5}$$

The vibrational transmission function in NEMD simulations is length-dependent and can be phenomenologically considered as (*33*, *49–51*)

$$T(\omega) = \frac{T_b(\omega)}{1 + L/\Lambda(\omega)} \tag{6}$$

where $T_b(\omega)$ denotes the phonon transmission function in the ballistic transport situation, $L$ is the transportation length, and $\Lambda(\omega)$ is phenomenologically frequency-dependent mean free paths of vibrations.

We also calculate the spectral energy density (SED) of the MHHPs through

$$\Phi(\vec{k}) = \frac{1}{4\pi\tau_0 N} \sum_\alpha^B \sum_b m_b \left| \int_0^{\tau_0} \sum_{n_{x,y,z}}^N v_\alpha \binom{n_{x,y,z}}{b}; t \times \exp\left[i\vec{k} \cdot \vec{r}\binom{n_{x,y,z}}{0} - i\omega t\right] dt \right|^2 \tag{7}$$

where $\tau_0$ is integration time which should be long enough, $N$ is the total number of unit cells, $m_b$ is the mass of the basic atom $b$ in the unit cell, $v_\alpha$ is atomic velocity along the $\alpha$ direction, $n_{x,y,z}$ is the index number of unit cells along $x$, $y$ and $z$ directions. The systems with a size of 8×8×40-unit cells were run in an *NVE* ensemble for 1 million steps, in which the atomic position and velocity were recorded. For MAPb(Br$_{0.75}$I$_{0.25}$)$_3$, MAPb(Br$_{0.5}$I$_{0.5}$)$_3$, and MAPb(Br$_{0.25}$I$_{0.75}$)$_3$, the blended halide atoms were assumed to retain the original crystalline sites.

## Acknowledgments

Y.Z. thanks the Equipment Competition fund (REC20EGR14) and the open fund from the State Key Laboratory of Clean Energy Utilization (ZJUCEU2022009) and the ASPIRE Seed Fund (ASPIRE2022#1) from the ASPIRE League. Z.L., Z.W. and Y.Z. acknowledge the fund from Research Grants Council of the Hong Kong Special Administrative Region under Grant C6020-22G. Y.Z. thanks for Research Grants Council of the Hong Kong Special Administrative Region under Grant 260206023. Y.Z. also thanks for the Hong Kong SciTech Pioneers Award from the Y-LOT Foundation. Y.G. thanks the fund from National Natural Science Foundation of China under Grant No. 52176166. The authors are grateful to the Materials Characterization and Preparation Facility (MCPF) of HKUST for their assistance in experimental characterizations.

## Author contributions

Y.Z. and G. W. conceived the idea; H.L. and Y.Z. supervised the project; G.W. designed the experiments and conducted the material synthesis, characterization, and performance investigation; H.F. and Y.G. did the calculations; Z.C. prepared the samples; G.W., F.H. and Y.Z. prepared the manuscript.; All the authors reviewed and revised the manuscript.

## Supplementary information

The online version contains supplementary information is available

25